\documentclass[pra,twocolumn
]{revtex4}

\usepackage{graphicx}
\usepackage{amsmath,hyperref}

\newcounter{mycount}

\newcommand{\half}{\frac{1}{2}}

\newcommand{\id}{\mathbf{1}}

\newcommand{\be}[1]{ \begin{eqnarray} \mbox{$\label{#1}$} }    
\newcommand{\ee}{\end{eqnarray}}

\usepackage{braket}
\usepackage{amsfonts,bm}
\DeclareMathOperator{\Tr}{Tr}

\begin{document}

\title{Improved Classical and Quantum Random Access Codes}
\author{O. Liab\o tr\o}
\affiliation{Department of Physics, University of Oslo, P.O. Box 1048 Blindern, 0316 Oslo, Norway}

\date{\today}

\begin{abstract} 
A (Quantum) Random Access Code ((Q)RAC) is a scheme that encodes $n$ bits into $m$ (qu)bits such that any of the $n$ bits can be recovered with a worst case probability $p>\frac{1}{2}$.  Such a code is denoted by the triple $(n,m,p)$.  It is known that $n<4^m$ for all QRACs and $n<2^m$ for classical RACs.  These bounds are also known to be tight, as explicit constructions exist for $n=4^m-1$ and $n=2^m-1$ for quantum and classical codes respectively.  We generalize (Q)RACs to a scheme encoding $n$ $d$-levels into $m$ (qu)-$d$-levels such that any $d$-level can be recovered with the  probability for every wrong outcome value being less than $\frac{1}{d}$.  We construct explicit solutions for all $n\leq \frac{d^{2m}-1}{d-1}$.  For $d=2$, the constructions coincide with those previously known.  We show that the (Q)RACs are $d$-parity-oblivious, generalizing ordinary parity-obliviousness.  We further investigate optimization of the success probabilities. For $d=2$, we use the measure operators of the previously best known solutions, but improve the encoding states to give a higher success probability.  We conjecture that for maximal $(n=4^m-1,m,p)$ QRACs, $p=\frac{1+\frac{1}{(\sqrt{3}+1)^m-1}}{2}$ is possible and show that it is an upper bound for the measure operators that we use.  When we compare $(n,m,p_q)$ QRACs with classical $(n,2m,p_c)$ RACs, we can always find $p_q\geq p_c$, but the classical code gives information about every input bit simultaneously, while the QRAC only gives information about a subset.  For several different $(n,2,p)$ QRACs, we see the same trade-off, as the best $p$ values are obtained when the number of bits that can be obtained simultaneously is as small as possible.  The trade-off is connected to parity-obliviousness, since high certainty information about several bits can be used to calculate probabilities for parities of subsets.


\end{abstract}
\pacs{ }

\maketitle


\section{Introduction}
\label{sec:intro}

The fundamental limits for how information can be encoded into a physical system and then retreived again lie at the core of quantum information theory.  Due to the Holevo bound\cite{Holevo}, n qubits can not transfer more than n classical bits of information faithfully.  However, interesting possibilities arise if we allow a limited chance for transmitting the wrong message.  Quantum Random Access Codes, (QRACs) exploit this. An $(n,m,p)$ QRAC encodes $n$ bits into $m$ qubits, such that any one bit can be retrieved with a worst case success probability $p>\half$.  The original QRACs\cite{tashma} include the (2,1,0.85) and (3,1,0.79) QRACs.  These QRACs were experimentally realized in 2009\cite{pcppom}.  It has been shown that $(4^m, m, p)$ QRACs are impossible\cite{impossible}, and that $(n,m,p)$ QRACs are possible for all $n<4^m$\cite{possible}.
Much of the research has also consentrated on maximising the average success probability.  If the communicating parties have access to shared randomness, the average success probability effectively becomes the worst case probability\cite{sr}.  Shared entangled states allow even more effective entanglement assisted random access codes\cite{earac}.  Entangled pairs also allow the super dense coding protocol\cite{sdc} and quantum teleportation\cite{qt}, where a qubit is used to send two bits faithfully in the first case and the other way around in the latter.  

We will neither consider shared randomness nor shared entanglement in this paper, but stick with the original idea of a (Q)RAC.   We use inherently parity-oblivious constructions and seek to optimize the worst case success probability for all $n$ that are possible in an $(n,m,p)$ (Q)RAC. We provide an explicit construction of QRACs for all $n<4^m$ which can also be employed for their classical counterparts, RACs, for all $n<2^m$.  The construction for QRACs was discovered in \cite{possible}, but we improve upon it by using better encoding states.  We generalize the problem to $d$-level (quantum) systems, encoding $n$ $d$-levels in such a way that every wrong outcome has a probability less than $\frac{1}{d}$.  The constructions used for $2$-levels are generalized to answer also this problem.  

This paper has the following structure:  We first give a proper definition of (Q)RACs.  Then we present a na\"ive numerical approach to find  $(n, 2, p)$ QRACs for n up to 12.  This approach uses only pure encoding states and projection-valued measures.  A more general approach uses mixed states based on the understanding of the geometry of density matrices.  We review this geometrical interpretation in section \ref{sec:geo}.  We then use this picture throughout the text to derive very general (Q)RACs.  The classical RACs can be seen as QRACs with diagonal density matrices.  We discuss the optimality of the derived codes and show that they are parity-oblivious in section \ref{sec:opt}.  The similarities and differences between $(n,m,p_q)$ QRACs and $(n,2m,p_c)$ RACs is the topic of section \ref{sec:qvsc}.     

\section{(Q)RACs}
An $(n,m,p)$ (Q)RAC consists of two parts, an encoding scheme, and a set of measurements.  The encoding scheme $e$ can be viewed as a function that takes a bit string $a$ of $n$ bits as input and returns a (quantum) state $\hat{\rho}_a$.
\be{}
a\overset{e}{\rightarrow} \hat{\rho}_a.
\ee
The input string $a$ will be represented by a binary number $0\leq a<2^n$ and the $i$'th digit of $a$ is the $i$'th input bit.  The (quantum) state $\hat{\rho}_a$ describes a physical state in a system of $m$ (qu)bits.  
For every bit of $a$, the (Q)RAC specifies a measurement $f_i$ that can be performed on the state $\hat{\rho}_a$, measuring the value of the $i$'th bit of $a$ to be $a_i'$:
\be{}
\hat{\rho}_a\overset{f_i\hspace{3pt}}{\rightarrow}a'_i.
\ee
The measurement $f_i$ is probabilistic and will not always give the correct bit value.  For the (Q)RAC to be valid, we require that
\be{pai}
p(a_i=a'_i)\geq p=\frac{1+\alpha}{2}>\half
\ee
for all input strings $a$ and all bit positions $i$.  We will sometimes refer to the value $\alpha$ in Eq. (\ref{pai}) as the bias of $p$.  We will assume that a standard basis is agreed upon and we represent operators acting on the physical system by matrices.  The communication between the canonical participants Alice and Bob performing a (Q)RAC can now be described by the chain
\be{}
a\overset{e}{\rightarrow} \rho_a\overset{f_i\hspace{3pt}}{\rightarrow} a_i'.
\ee
Alice obtains the string $a$, encodes it in the physical system described by the density matrix $\rho_a$ and sends the system to Bob who performs a measurement and obtains the correct value for the $i$'th bit with a probability at least $p$. 

\section{Pure state $(n,2,p)$ QRACs}
\label{sec:pure}

The original and optimal $(n,1,p)$ QRACs use only pure states\cite{tashma}.  Mixed states become useful for quantum systems of dimension higher than two.  It is interesting all the same to investigate how large n can be if only pure encoding states are allowed.  To achieve this, we have performed numerical searches focusing on $(n,2,p)$ QRACs.

A $(7,2,0.54)$ QRAC was proposed in \cite{impossible}.  We find that n can be at least 12.  It is still an open question if 13, 14 or even 15 is possible.

We have performed numerical searches with the following setup: 

\be{}
B=\{B_j\ |\ j\in\{0,...,n-1\}\}
\ee
is a set of n orthonormal bases for $\mathbb{C}^4$, representing the Hilbert space of $m=2$ qubits.  The $j$'th basis defines the measurement of the j'th bit such that projecting onto one of the two first basis vectors corresponds to the bit being 0 and the other two to the bit being 1.  

Further, we define 
\be{}
R=\{\ket{a}\ |\ i\in\{0,...,2^n-1\}\}
\ee
as the encoding states.  The encoding states must be chosen such that if the $j$'th bit/digit of $a$ is $0(1)$, then the projection of $\ket{a}$ onto the two first(last) basis vectors of $B_j$ has absolute square greater than $\half$.

With a fixed set of bases, any randomly drawn state vector will encode an input string with a worst case probability greater than $\half$ unless both outcome probabilities for a measurement are exactly $\half$.  This is however highly unlikely when drawing random state vectors.  One possible approach is therefore to draw random bases and then draw random state vectors until, hopefully, all input states are represented by a state vector.  This approach will not give optimal QRACs, but will, if successful, show that a QRAC is possible for a given n.  We find that QRACs up to n=9 are found within minutes on a desktop computer with this method when drawing $\approx$ $10^2$ random bases and $10^6$ state vectors for each basis. This method also quickly found an $n=7$ QRAC with worst case $p>0.58$, which improves the result in \cite{impossible}.

We may make several improvements to this approach.  Firstly, we may search for optimal encoding states for each input using a random walk algorithm.  Secondly, we may also make small, random adjustments to the measurement bases, and keep the changes if the worst case $p$ is improved after a round of optimizing encoding states.  Finally, we may set some conditions on the initial measurement bases.   Each measurement basis consists of two planes that correspond to $0$ and $1$ respectively.  If one such plane has a very small overlap with another plane, then the combination of bits that they represent will be less likely.  Therefore, we want the planes to exhibit a certain degree of mutual unbiasedness.  One way to ensure this is to draw random starting bases until the worst case overlap between planes is over a treshold value.

The state vectors and measurement bases obtained using numerical searches are available on \url{http://folk.uio.no/olalia/QRAC}.  The average and worst case probabilities are listed in table \ref{table:1}.
\begin{table}

\begin{center}
\begin{tabular}{|l|c|c|c|c|c|c|}
\hline
$n$ & 7 & 8 & 9 & 10 & 11 & 12\\ \hline
$p$ & 0.68412  & 0.65249  &  0.60319  & 0.53919  & 0.52468 & 0.50054 \\ \hline
$\bar{p}$ & 0.72839 & 0.71653  & 0.70268 & 0.66544  & 0.66177 & 0.65562\\ \hline
\end{tabular}
\caption{\label{table:1} Numerically obtained $(n,2,p)$ QRACs.  $\bar{p}$ is the average success probability over all input states and measured bits.}
\end{center}
\end{table}

\section{Density matrix geometry}
\label{sec:geo}

General encoding states are described by density matrices, and these can be understood geometrically in terms of their Bloch vectors.  We will now briefly review this.  For more details, see refs. \cite{Jakobzyk, apophatic}.

The density matrix of a qubit in a pure state can be expressed as 

\be{}
\rho=\frac{1}{2}(\id+\bm{r}\cdot\bm{\sigma})
\ee
where $\bm{\sigma}=(\sigma_x, \sigma_y, \sigma_z)$ are the Pauli matrices, obeying $\{\sigma_{k}, \sigma_{k'}\}=2\delta_{kk'}$, and the Bloch vector $\bm{r}\in \mathbb{R}^3$ with $|\bm{r}|=1$.  This gives a one to one correspondence between the Bloch sphere and pure states in a two-level system.  Mixed states are weighted averages of pure states, and they fill the sphere, creating the Bloch ball.  Thus a general density matrix has a Bloch vector obeying $|\bm{r}|\leq 1$.  If $\rho$ and $\rho'$ are two density matrices, with corresponding Bloch vectors $\bm{r}$ and $\bm{r}'$, then the expectation value for the overlap can be expressed in terms of the Bloch vectors as

\be{}
\Tr(\rho\rho')=\half(1+\bm{r}\cdot\bm{r}')
\ee

The Bloch vectors can be generalized to N-level systems, and describe the set of $N\times N$ density matrices.  We denote the set of such Bloch vectors by $\mathcal{B}_N\subset \mathbb{R}^{N^2-1}$. The Bloch vectors of pure states are a part of this set with topology $\mathbf{CP}^{N-1}$.  We define $\bm{\sigma}\equiv(\sigma_1,...,\sigma_{N^2-1})$ as the generalized $N\times N$ Gell-Mann matrices, all obeying 
\be{orthogonalgenerators}
\Tr(\sigma_k\sigma_{k'})=2\delta_{kk'}.
\ee  
They span the set of traceless hermitian matrices, so if $\rho_{\bm{\alpha}}$ is a density matrix, then there is a vector $\bm{\alpha}\in\mathbb{R}^{N^2-1}$ such that
\be{}
\rho_{\bm{\alpha}}=\frac{1}{N}\id + \half\sum_{k=1}^{N^2-1}\alpha_k\sigma_k
\ee
The converse is however not true. A density matrix must have non-negative eigenvalues, and only the convex subset $\mathcal{B}_N$ of $\mathbb{R}^{N^2-1}$ corresponds to density matrices.

The condition (\ref{orthogonalgenerators}) gives the overlap of two density matrices the simple form
\be{overlap}
\Tr(\rho_{\bm{\alpha}}\rho_{\bm{\beta}})=\frac{1}{N}+\frac{1}{2}\bm{\alpha}\cdot\bm{\beta}.
\ee
It follows that 
\be{}
\Tr(\rho_{\bm{\alpha}}^2)=\frac{1}{N}+\half|\bm{\alpha}|^2=\sum_{k=1}^N p_k^2,
\ee
where $p_k$ are the eigenvalues of $\rho_{\bm{\alpha}}$.  This gives the length of a Bloch vector, 
\be{length}
|\bm{\alpha}|=\sqrt{2(-\frac{1}{N}+\sum_{k=1}^N p_k^2)}.
\ee
The length is maximized if exactly one eigenvalue is non-zero, or equivalently the density matrix corresponds to a pure state.  This gives the radius of the outsphere of the Bloch space,
\be{}
R_N=\sqrt{\frac{2(N-1)}{N}}.
\ee
All the pure states lie on this sphere, but they only make a $2(N-1)$ dimensional subspace of the $N^2-2$-dimensional sphere.  For $N=2$, they coincide, but for $N>2$, the outsphere is mainly invalid Bloch vectors.  The radius of the insphere also follows from equation (\ref{length}).  The insphere is the largest sphere that is contained in the Bloch space.  The radius is given by the smallest length of any Bloch vector on the surface of $\mathcal{B}_N$.  A density matrix has a surface Bloch vector iff it has at least one eigenvalue which is zero, making it infinitesimally close to a non-non-negative matrix.  If we assume that at least one eigenvalue is 0, then (\ref{length}) is minimized when all other eigenvalues are $\frac{1}{N-1}$.  This gives the radius of the insphere, 
\be{}
r_N=\sqrt{\frac{2}{N(N-1)}}.
\ee  
Any Bloch vector with radius at most $r_N$ corresponds to a density matrix.

It follows from non-negativity of overlaps that 
\be{}
cos(\angle(\alpha,\beta))\geq\frac{-1}{N-1}
\ee
if $\alpha$ and $\beta$ correspond to pure states\cite{Jakobzyk}.  Equality occurs when the state vectors are orthogonal.  This also means that an orthogonal basis corresponds to the corners of a simplex in Bloch space. 

When two Bloch vectors are orthogonal, Eq. (\ref{overlap}) shows that the overlap of the density matrices is $\frac{1}{N}$. The density matrices are said to be mutually unbiased in this case.  Likewise, we say that two state vectors are mutually unbiased if they have mutually unbiased density matrices.  Two orthonormal bases for $\mathbb{C}^N$ are mutually unbiased if the basis vectors of one of the bases are mutually unbiased with the basis vectors of the other.  A set of bases where all pairs of bases are mutually unbiased is a set of mutually unbiased bases (MUB).  In terms of Bloch space, MUB are vertices of simplexes lying in perpendicular subspaces.  Since an N-simplex is an $N-1$ dimensional object and the Bloch space has dimension $N^2-1$, the maximal number of mutually unbiased bases is at most $\frac{N^2-1}{N-1}=N+1$.  It is known\cite{mub} that for powers of a prime N, $N+1$ MUB can be constructed.  Surprisingly, it is not known for any composite numbers of distinct primes if this is the case or not. Only three MUB have been found in dimension 6, but it is not even proven that the maximal number is less than 7\cite{mub}.

We round off this section with a remark on quantum measurements.  A positive operator-valued measure (POVM) is a set of operators with non-negative eigenvalues that sum up to the identity.  In this article, we will be interested in a set of $n$ such measures and each measure will have $d$ different outcomes.  The $i$'th measure is denoted by 
\be{}
F_i=\{\hat{F}_{ij}|j\in\{0,...,d-1\}\}.
\ee
Performing the $i$'th measurement on a state $\hat{\rho}$ gives the probability
\be{}
p_{ij}=\Tr(\hat{F}_{ij}\hat{\rho})
\ee
for the $j$'th outcome. The operators $\hat{F}_{ij}$ do not in general have unit trace, but since they have non-negative eigenvalues, they are proportional to density operators, and we may use Eq. (\ref{overlap}) to calculate probabilities.  We will associate a measure operator with the Bloch vector 
\be{}
\bm{\beta}(\hat{F})=\bm{\beta}\ :\ \frac{1}{\Tr(\hat{F})}\hat{F}=\hat{\rho}_{\bm{\beta}}
\ee
and the overlap with a state $\hat{\rho}_{\bm{\alpha}}$ is then 
\be{}
\Tr(\hat{F}\rho_{\bm{\alpha}})=\Tr(\hat{F})(\frac{1}{N}+\half\bm{\alpha}\cdot\bm{\beta}(\hat{F})).
\ee

Projection-valued measures (PVMs) are a special case of POVM.  Their operators have eigenvalues that are all either $0$ or $1$.  Since they are projection operators, we will denote them by
\be{}
\Pi_i=\{\hat{\pi}_{ij}|j\in\{0,...,d-1\}\}.
\ee
Now, if $\Tr(\hat{F})=\Tr(\hat{\pi})\in\mathbb{Z}$, then (\ref{length}) implies that 
\be{}
|\bm{\beta}(\hat{F})|\leq |\bm{\beta}(\hat{\pi})|
\ee
where equality only occurs if $\hat{F}$ is in fact a PVM operator.  Because of this, we will prefer PVMs, whenever we can find them with the Bloch vector directions that we need.

\section{$(n,m,p)$ QRACs}
\label{sec:two}

We now explicitly construct $(n, m,p)$ QRACs for all $n<4^m$.  A $(4^m-1,m,\frac{1+\frac{1}{(2^m-1)\sqrt{2^m+1}}}{2})$ QRAC was demonstrated in \cite{possible}.  This construction can also be used for all  $n<4^m$ and generalizes the original $(n,1,p)$ QRACs, but utilizes mixed states in order to place the encoding states on a hypercube.  We give a detailed description of this solution and adjusts it to the more general $n<4^m$.  We first give a solution with both encoding states and measure operators based on the insphere of Bloch space.  Subsequently, we make improvements and arrive at the solution found in \cite{possible}.  We improve the solution further in section \ref{sec:opt}\\

\subsection{Insphere-based solution}\label{insphere}

The dimension of the Hilbert space for $m$ qubits is $N=2^m$.  Let $\bm{\sigma}=\sum_{k=1}^{N^2-1}\sigma_k\bm{e}_k$ be the generalized Gell-Mann matrices and define $n$, up to $N^2-1=4^m-1$ POVM matrices with $d=2$ different outcomes as  
\be{gensolmeasure}
F_{ij}=\frac{N}{2}\rho_{(-1)^jr_N\bm{e}_i}=\half\id+(-1)^j\half\sqrt{\frac{N}{2(N-1)}}\sigma_i.
\ee

The bit strings that are encoded can be identified with functions, $\beta : \{1,..,n\}\rightarrow \{0,1\}$.  We now define the encoding Bloch vectors for each such string, 
\be{}
\bm{\beta}=\sum_{k=1}^{n}(-1)^{\beta(k)}\sqrt{\frac{2}{N(N-1)}}\frac{1}{\sqrt{n}}\bm{e}_k ,
\ee
with corresponding density matrices $\rho_{\bm{\beta}}$.  The Bloch vector length of $r_N$ ensures that all the states are valid.  If we perform the $i$'th measurement, we get the probability
\be{}
p_{ij}=\Tr(\rho_{\bm{\beta}}F_{ij})=\frac{1+(-1)^{\beta(i)+j}\frac{1}{(N-1)\sqrt{n}}}{2}
\ee
for the outcome $j$.  If $\beta(i)=j$, then the result $j$ has probability $p_{ij}>\half$.  We conclude that the POVM $F_i$ determines the value of the $i'th$ bit with a success probability

\be{}
p=\frac{1+\frac{1}{(N-1)\sqrt{n}}}{2}=\frac{1+\frac{1}{(2^m-1)\sqrt{n}}}{2}.
\ee
We see that this reproduces the well known optimal results for $n=2,3$ and $m=1$.  This is because the insphere of the Bloch sphere is the Bloch sphere itself.  The success probability when encoding the maximal number of $n=4^m-1$ bits is 
\be{}
p(n=4^m-1)=\frac{1+\frac{1}{(2^m-1)\sqrt{4^m-1}}}{2}.
\ee

\subsection{Improved QRACs}
\label{subsec:good_two}

We now improve the insphere based solutions, reaching the known solution arrived at in \cite{possible}.  In order to improve the solution, we must see if either the Bloch vectors of the encoding states or those of the POVMs have non-maximal length and may be scaled up to the surface of $\mathcal{B}_N$.  Since we have only $2n$ POVM operators, in contrast to the $2^n$ encoding states, it may seem easiest to do this with the POVMs.

The POVM operators of the previous subsection were proportional to density matrices with Bloch vectors in the same directions as the generalized Gell-Mann matrices.  The requirement that 
\be{}
F_{ij}=\frac{N}{2}(\frac{1}{N}\id +(-1)^j\half|\bm{\alpha}|\sigma_i)
\ee
has non-negative eigenvalues gives the restriction
\be{}
|\bm{\alpha}|\leq\frac{2}{N\cdot \mbox{max}_{p_i \in \mbox{eigenvalues}(\sigma_k)}(|p_k|)}\ \forall \sigma_k
\ee
The generalized Gell-Mann matrices includes a matrix with an eigenvalue of $-\sqrt{\frac{2(N-1)}{N}}$ and this gives a worst case Bloch vector length for the measurement operators of $r_N$.  It is however possible to use matrices with different eigenvalues, as long as they are linearly independent, traceless and fulfill $\Tr(\sigma_k\sigma_{k'})=2\delta_{kk'}$.  We will now define such matrices.

Let $\tilde\sigma_0=\id_2$, let $\tilde\sigma_1=\sigma_x,\ \tilde\sigma_2=\sigma_y,\ \tilde\sigma_3=\sigma_z$ be the Pauli matrices, and let $c_i(k_4)$ be the i'th digit from the right in the representation of an integer $k$ in base 4.  An alternative set of matrices is then given by the set
\be{altmatrix}
\{\sigma_k=2^{\frac{1-m}{2}}\bigotimes_{i=1}^m \tilde\sigma_{c_i(k_4)}\}_{k=1}^{4^m-1}.
\ee
It is straightforward to check that they obey the requirement $\Tr(\sigma_k\sigma_{k'})=2\delta_{kk'}$.  The eigenvalues of each matrix are $2^{m-1}$-fold degenerate with the values $\pm 2^{\frac{1-m}{2}}$.  This gives the restriction 
\be{}
|\bm{\alpha}|\leq 2^{\frac{1-m}{2}}=\sqrt{\frac{2}{N}}
\ee
and we may choose equality here when constructing a QRAC.  Doing so allows us to improve the measurement operators of Eq. (\ref{gensolmeasure}) to
\be{goodsolmeasure}
\pi_{ij}=\frac{N}{2}\rho_{(-1)^j\sqrt{\frac{2}{N}}\bm{e}_i}=\half\id\pm(-1)^j\sqrt{\frac{N}{2}}\sigma_i.
\ee
The measure operators have eigenvalues that are all either 0 or 1, so the measurements are PVMs.  Choosing encoding states on the insphere as in the previous solution gives the success probability

\be{}
p(m, n<4^m)=\frac{1+\frac{1}{\sqrt{(2^m-1)n}}}{2}.
\ee
For maximal $n=4^m-1$, we have

\be{}
p=\frac{1+\frac{1}{(2^m-1)\sqrt{2^m+1}}}{2}.
\ee
The improved success probabilities are shown in table \ref{goodnumvals}.

\begin{table}
\begin{center}
\begin{tabular}{|c|c|c|c|}
\hline
$m$ & $n$ & exact $p$ & $p$ \\ \hline 
1 & 3 & $\half(1+\frac{1}{\sqrt{3}})$ & .78868\\ \hline 
2 & 15 & $\half(1+\frac{1}{3\sqrt{5}})$ & .57454\\ \hline
3 & 63 & $\half(1+\frac{1}{21})$ & .52381\\ \hline
4 & 255 & $\half(1+\frac{1}{15\sqrt{17}})$ & .50808\\ \hline
5 & 1023 & $\half(1+\frac{1}{31\sqrt{33}})$ & .50281\\ \hline
\end{tabular}
\caption{\label{goodnumvals} Improved success probabilities for QRACs encoding $n=4^m-1$ bits into $m$ qubits.}
\end{center}
\end{table}

Each orthogonal measurement in the original $(n,1,p)$ QRACs did only give information about one chosen bit to be measured.  This is not the case for $m>1$.  If one is interested in bit number $i$, then one makes a measurement in the $\sigma_i$ eigenbasis, but different $\sigma_i$ may have common eigenbases.  For example, if we have 
\be{sigmaxyz}
\sigma_i=\sigma_x\otimes\sigma_y\otimes\sigma_z,
\ee
then the tensor products of the eigenbases for the Pauli matrices make an eigenbasis for $\sigma_i$.  But this basis is also an eigenbasis for any of the six other $\sigma_{i'}$ that one can obtain by substituting any, but not all of the Pauli matrices in (\ref{sigmaxyz}) with $\id_2$.  If we make a measurement in this basis, we can interpret it for any of 7 different bits.  The probabilities we get then are however not independent.  We will look at the joint probabilities in section \ref{sec:po}.

\section{$(2^{m}-1,m,p)$ RACs}
\label{sec:ctwo}

We now demonstrate that an $(n=2^m-1,m,\frac{1+\frac{1}{n}}{2})$ RAC where m classical bits are transmitted is possible.  This is done using purely classical, local randomness.  A set of classical bits is equivalent to qubits if we demand the density matrices to be diagonal.  This gives a simplex-shaped Bloch space, with density matrices on the form 

\be{}
\rho_{\bm{\alpha}}=2^{-m}\id+\frac{1}{2}\sum_{i=1}^{2^m-1}\alpha_i\sigma_i 
\ee
where the $\sigma$ matrices are limited to the $2^m-1$ diagonal matrices of the set (\ref{altmatrix}).  They are tensor products of $\id_2$ and $\sigma_z$.  The measurement operators and encoding density matrices are chosen in the same way as for the improved QRACs and we get the same worst case probability as for QRACs, but with a smaller allowed n.

\be{}
p(m, n<2^{m})=\frac{1+\frac{1}{\sqrt{(2^m-1)n}}}{2}.
\ee
In the maximal case $n=2^m-1$, this reduces to
\be{}
p=\frac{1+\frac{1}{n}}{2}.
\ee
We will see later that this success probability can also be obtained for non-maximal n.  Comparing tables \ref{goodnumvals} and \ref{classicnumvals}, we see that the QRAC encodes $2^m+1$ times as many bits as the RAC, at the cost of a factor $\frac{1}{\sqrt{2^m-1}}$ in the bias.

\begin{table}
\begin{center}
\begin{tabular}{|c|c|c|c|}
\hline
$m$ & $n$ & exact $p$ & $p$ \\ \hline 
1 & 1 & 1 & 1\\ \hline 
2 & 3 & $\half(1+\frac{1}{3})$ & .66667\\ \hline
3 & 7 & $\half(1+\frac{1}{7})$ & .57143\\ \hline
4 & 15 & $\half(1+\frac{1}{15})$ & .53333\\ \hline
5 & 31 & $\half(1+\frac{1}{31})$ & .51613\\ \hline
\end{tabular}
\caption{\label{classicnumvals} Success probabilities for RACs encoding $n=2^{m}-1$ bits into $m$ bits.}
\end{center}
\end{table}

The $(3,2,p)$ case can be explained in simple terms without density matrices.
Alice gets to know the bit string and encodes it in a RAC that she sends to Bob.  They have agreed beforehand that Alice will send two bits that signify the values on the two first input bits.  If Bob wants the third bit, then he will assume that the bit sum of all the input bits is 0 modulo 2.  If the input bits have 0 sum modulo 2, then Alice will just send the two first bits and Bob will make the correct assumption.  If the bit sum is odd, then Alice will use a randomness generator that sends one of three messages, each with probability $\frac{1}{3}$.  The messages are either the two first bits, the first bit flipped and the second bit unchanged or the first bit with the second bit flipped.  The three different messages make Bob guess wrong on one bit each, giving a worst case probability of $\frac{2}{3}$ for any given bit.  In terms of Bloch geometry, the corners of a tetrahedron corresponds to even bit sum inputs and the midpoints on the surfaces corresponds to odd bit sum inputs.  The measurement directions are the midpoints of the edges.  We have taken the liberty to improve the solution by placing the even bit sum inputs outside the insphere, giving an average probabilty of $\frac{5}{6}$ if the input is uniform, but the same worst case p.  The unimproved solution has a 50\% chance for flipping two bits if the bit sum is already even, making the success probability $\frac{2}{3}$ also in this case.

\section{Generalization to (quantum) d-levels}
\label{sec:d}
We will now generalize the above results to $d$-level systems.  Such QRACs have been considered before\cite{wigner, dlevel}, but only for maximising the average success rate with uniform input.  We need to define our generalization of the worst case problem to $d$-level QRACs.  For $d=2$, the worst case probability is required to be larger than $\half$.  The natural generalization is to demand 
\be{weak}
p>\frac{1}{d}
\ee
 as this is the limit for beating a pure guess.  However, this is not the only possible generalization of the problem.  For $d=2$, we also have that the probability for \textit{failure} is \textit{less} than $\half$.  We may generalize this to the stronger requirement
\be{strong}
p_j<\frac{1}{d} \forall j\neq \tilde{j}
\ee
where $\tilde{j}$ is the correct outcome.  This of course also implies the weaker condition (\ref{weak}).  We will choose this last requirement for our $d$-level QRACs.  If we had only required the weaker condition, then a $d'$-level QRAC could be made into a $d$-level QRAC with $d>d'$ by dividing the $d$ outcomes into $d'$ groups of outcomes and then make a guess inside the group.  For example, a classical bit can be used to guess an octet with probability $p=\frac{1}{4}$.  With the stronger condition (\ref{strong}), the angle between a wrong encoding state and a measurement direction must be obtuse ($>90^\circ$).  In order to fit $d$ vectors with obtuse angles between them, we need a $d-1$ dimensional state space, so for classical RACs, the transmitted system must be at least a $d$-level, while a QRAC needs at least a qu-$\sqrt{d}$-level.  With the strong condition, the natural optimization problem would be to minimize the largest wrong outcome probability.  However, our constructions make sure that all the wrong outcomes have the same probability, and we can therefore maximize the worst case success probability as before.

\subsection{Insphere method}
We generalize the insphere method of section \ref{insphere}.  The measurement basis improvement that was possible for $d=2$ is possible if $d$ is a power of a prime, which we will show in the next section.\\

The mixed states of m $d$-levels can be described by the Bloch vectors $\mathcal{B}_{d^m}$.  The insphere has a radius

\be{}
r_{d^m}=\sqrt{\frac{2}{d^m(d^m-1)}}
\ee
and is the boundary of a solid $d^{2m}-1$ dimensional ball.  To fulfill (\ref{strong}), we need a $d-1$ dimensional subspace for each POVM.   We choose n, up to $\frac{d^{2m}-1}{d-1}$, hyperplanes of dimension $d-1$.  We now pick vertices of a simplex on the intersection of the $i$'th plane and the insphere and name the $j$'th vertex $\bm{\alpha}_{ij}$.  This gives POVM matrices

\be{}
F_{ij}=d^{m-1}\rho_{\bm{\alpha}_{ij}}.
\ee

We define the encoding states
\be{}
\rho_a=\rho_{\bm{\beta}_a},\ \bm{\beta}_a=\frac{r_{d^m}}{\sqrt{n}}\sum_{i=1}^n \bm{\alpha}_{ic_i(a_d)}.
\ee
where $c_i(a_d)$ denotes the $i$'th digit of $a$ in base $d$.  The success probability follows from Eq. (\ref{overlap})

\be{}
p=\frac{1+\frac{1}{(d^m-1)\sqrt{n}}}{d}.
\ee
For the  maximal $n=\frac{d^{2m}-1}{d-1}$ QRACs we have

\be{}
p=\frac{1+\frac{\sqrt{d-1}}{(d^m-1)\sqrt{d^{2m}-1}}}{d}.
\ee

We can also construct classical RACs using the insphere of a solid $d^m$ simplex.  The success probability will be the same, but the maximal number of input $d$-levels is $\frac{d^m-1}{d-1}$.

\subsection{Dimension being a power of a prime}
\label{sec:powerofprime}

We will now make an improved version of the QRAC presented in the previous section.  It is guaranteed to work if $d$ is the power of a prime due to two fascinating results.  We will start by considering classical RACs.  The generalization to QRACs is straightforward when using mutually unbiased bases.\\

Ideally, we want all the measure operators to have eigenvalues that are either 0 or 1, that is, they are PVM operators.  A PVM is then described by a set of $d$ diagonal matrices, each with $d^{m-1}$ $1$'s on the diagonal.  Two PVM's should also be mutually unbiased, meaning that if $i\neq i'$, we have
\be{mupvm}
\Tr(\pi_{ij}\pi_{i'j'})=d^{m-2}\ \forall\ j,j'\in\{0,...,d-1\}.
\ee
This makes sure that the associated Bloch vectors of two PVM's span orthogonal subspaces.  For $d=2$, $m=2$, two mutually unbiased PVMs are

\be{}
\pi_{00}=\left(\begin{array}{cccc}
1 & 0 & 0 & 0\\
0 & 1 & 0 & 0\\
0 & 0 & 0 & 0\\
0 & 0 & 0 & 0\end{array}\right),\ \pi_{01}=\left(\begin{array}{cccc}
0 & 0 & 0 & 0\\
0 & 0 & 0 & 0\\
0 & 0 & 1 & 0\\
0 & 0 & 0 & 1\end{array}\right)
\ee
and
\be{}
\pi_{10}=\left(\begin{array}{cccc}
1 & 0 & 0 & 0\\
0 & 0 & 0 & 0\\
0 & 0 & 1 & 0\\
0 & 0 & 0 & 0\end{array}\right),\ \pi_{11}=\left(\begin{array}{cccc}
0 & 0 & 0 & 0\\
0 & 1 & 0 & 0\\
0 & 0 & 0 & 0\\
0 & 0 & 0 & 1\end{array}\right)
\ee
and one more mutually unbiased PVM is also possible.  Since the matrices are all diagonal, and since the matrices of a single PVM are non-overlapping, we may represent the PVM's in a compact matrix form.  We define the $n\ \times\ d^m$ matrix
\be{}
M_{ik}=\sum_{j=1}^d\pi_{ij,kk}\cdot j
\ee
where $\pi_{ij,kk}$ is the $k$'th diagonal element of the matrix $\pi_{ij}$.  The measurement operators can then be read from the matrix, 
\be{}
\pi_{ij,kk}=\delta_{M_{ik},j}.
\ee
As an example, for d=4, m=2, we have the matrix
\be{M4}
M=\left(\begin{array}{cccccccccccccccc}
0 & 0 & 0 & 0 & 1 & 1 & 1 & 1 & 2 & 2 & 2 & 2 & 3 & 3 & 3 & 3\\
0 & 1 & 2 & 3 & 0 & 1 & 2 & 3 & 0 & 1 & 2 & 3 & 0 & 1 & 2 & 3\\
0 & 1 & 2 & 3 & 2 & 3 & 0 & 1 & 3 & 2 & 1 & 0 & 1 & 0 & 3 & 2\\
0 & 1 & 2 & 3 & 3 & 2 & 1 & 0 & 1 & 0 & 3 & 2 & 2 & 3 & 0 & 1\\
0 & 1 & 2 & 3 & 1 & 0 & 3 & 2 & 2 & 3 & 0 & 1 & 3 & 2 & 1 & 0\end{array}\right).
\ee
Each row corresponds to a PVM, and the $k$'th element says which of the $d$ projection operator matrices that has a 1 in the $k$'th position on the diagonal.  The mutual unbiasedness of the PVM translates into the matrix property that for every pair of rows, every ordered pair of numbers from the same column occurs $d^{m-2}$ times.  Such a matrix is called an orthogonal array, and is equivalent to $n-m$ orthogonal latin hypercubes of dimension $m$\cite{latin}.  If $d$ is a power of a prime, then one can construct an orthogonal array with $\frac{d^{m}-1}{d-1}$ rows\cite{kishen}.  Such an array gives us $\frac{d^{m}-1}{d-1}$ mutually unbiased PVMs. This is the maximal number, since each PVM spans a $d-1$ dimensional subspace of the $d^m-1$-dimensional Bloch space and the subspace is orthogonal to the subspaces spanned by other PVMs.



We can now define a classical RAC.  We define the $a$'th encoding density matrix as  

\be{racenco}
\rho_a=\frac{1}{n}\sum_{i=1}^n d^{1-m}\pi_{ic_i(a_d)}
\ee
which gives a success probability
\be{cracd}
p_{ic_i(a_d)}=\Tr(\rho_a\pi_{ic_i(a_d)})=\frac{1+\frac{d-1}{n}}{d}.
\ee
We may improve some of the encoding states by extending their Bloch vectors to maximal size, but there will always be some encoding states that are already on the surface of the Bloch space and the worst case probability is therefore given by (\ref{cracd}).  An encoding state is on the surface if it has 0 as an eigenvalue.  If we read eigenvalues along the diagonal, then the $k$'th eigenvalue of $\rho_a$ will be 0 if $c_i(a_d)\neq M_{ik}\ \forall i$.  For example, with $M$ as in (\ref{M4}), the first eigenvalue of $\rho_a$ will be 0 iff none of the digits of $a$ are 0.\\

We can now go on to $d$-level QRACs.  We use the fact that there are $d^m+1$ MUB when $d$ is a power of a prime.  Then for each basis, we use the RAC construction to get mutually unbiased PVMs.  We then get a total of $(d^{m}+1)\frac{d^m-1}{d-1}=\frac{d^{2m}-1}{d-1}$ PVMs.  We define the PVMs as follows:

Let $\rho_{h,k}$ be the density matrix of the $k$'th basis vector of the $h$'th basis.  $M$ is the same $n$ by $d^m$ matrix as for the classical RAC.  The PVMs are defined by

\be{dlevelpvm}
\pi_{ij}=\sum_{k=1}^{d^m} \delta_{M_{i\%(d^m+1);k},j}\cdot\rho_{i//(d^m+1),k},
\ee 
where $//$ and $\%$ denotes integer division and modulo.  We have separated the indices of $M$ by a semicolon for clarity.  Each $\pi_{ij}$ is diagonal in one of the MUB in a way given by a row of the orthogonal array M. 

We can define the encoding states in the same way as for classical RACs (\ref{racenco}) and in this case, the success probability is the same as for RACs (\ref{cracd}).

Alternatively, we may place the encoding states on the insphere.  From the eigenvalues of $\pi_{ij}$ and Eq. (\ref{length}), we find that 
\be{}
\pi_{ij}=d^{m-1}\rho_{\bm{\alpha}_{ij}}, \ \ |\bm{\alpha}_{ij}|=\sqrt{2d^{-m}(d-1)}
\ee
where $\alpha_i$ is a Bloch vector.  Each encoding state of (\ref{racenco}) has a Bloch vector which is the average of $n$ orthogonal $\bm{\alpha}_{ij}$, and therefore has length 
\be{}
\rho_a=\rho_{\bm{\alpha}_a},\ \ |\bm{\alpha}_a|=\frac{1}{\sqrt{n}}\sqrt{2d^{-m}(d-1)}.
\ee
On the other hand, the insphere radius is 
\be{}
r_{d^m}=\sqrt{\frac{2}{d^m(d^m-1)}}.
\ee
The ratio is
\be{ratio}
\frac{r_{d^m}}{|\bm{\alpha}_a|}=\sqrt{\frac{n(d-1)}{d^m-1}}
\ee
and the success probability with encoding states on the insphere can be obtained from Eq. (\ref{cracd})

\be{pdinsph}
p=\frac{1+\frac{r_{d^m}}{|\bm{\alpha}_a|}\frac{d-1}{n}}{d}=\frac{1+\sqrt{\frac{d-1}{n(d^m-1)}}}{d}
\ee

The ratio in Eq. (\ref{ratio}) indicates which encoding states to use and we may merge (\ref{cracd}) and (\ref{pdinsph}) to obtain

\be{pbound}
p=\left\{\begin{array}{ll}
\frac{1+\sqrt{\frac{d-1}{n(d^m-1)}}}{d} & n\geq(d-1)(d^m-1)\\
\frac{1+\frac{d-1}{n}}{d} & n\leq (d-1)(d^m-1) \end{array}\right.
\ee

This summarizes the worst case success probability for our (Q)RAC constructions so far, with classical RACs for all $n\leq\frac{d^m-1}{d-1}$ and QRACs for all $n\leq \frac{d^{2m}-1}{d-1}$.  We have used maximal sets of mutually unbiased PVMs as well as encoding Bloch vectors with lengths that guarantee valid states.  It is however sometimes possible to use longer Bloch vectors.  We will now see that finding the encoding states that gives the best worst case probability can be formulated as an eigenvalue problem.\\

Eq. (\ref{dlevelpvm}) defines $\frac{d^{2m}-1}{d-1}$ mutually unbiased PVMs.  We assume that the QRAC uses a subset of n PVMs and redefine the indices such that they run from $0$ to $n-1$ for any choice of subset.  The encoding states can be defined as in eq. (\ref{racenco}), but with a scaling factor on the traceless part.
\begin{equation}
\rho_a=K(\frac{1}{n}\sum_{i=1}^n d^{1-m}\pi_{ic_i(a_d)}-d^{-m}\id)+d^{-m}\id
\end{equation} 
The Bloch vector length is proportional to the traceless part, so we have the new success probability
\begin{equation}
p=\frac{1+K\frac{d-1}{n}}{d}.
\end{equation}
We demand that all the eigenvalues of $\rho_a$ lie between $0$ and $1$, as it is a density matrix, and this gives
\begin{equation}
-1\leq \frac{K}{n}\text{eigenvalue}(\sum_{i=1}^{n}(d\pi_{ic_i(a_d)}-\id))\leq d^m-1.
\end{equation} 

Both inequalities must hold for all eigenvalues, but the first inequality implies the second, since the matrix is traceless.  We can therefore neglect the second and concentrate on the first.  We denote the most negative eigenvalue of all the $\sum_{i=1}^{n}(d\pi_{ic_i(a_d)}-\id)$ where $0\leq a<d^n$ by $-\lambda$.  We then have
\be{}
\frac{K}{n}\leq\frac{1}{\lambda}
\ee 
and we can write the worst case probability as 
\be{poflambda}
p=\frac{1+\frac{d-1}{\lambda}}{d}.
\ee
It is however not in general easy to find $\lambda$, we will consider this problem in section \ref{sec:opt}.

\section{Parity-obliviousness}
\label{sec:po}
Parity-obliviousness is a cryptographic property that pertains to some (Q)RACs.  If $S$ is a subset of input bits, then the parity of this set is the bit sum modulo 2.  A (Q)RAC is parity-oblivious iff no information can be obtained about the parity of any subset of at least two bits, when the input has been chosen randomly from the uniform distribution.

It is known that for $d=2$, $2n\leq m$, a parity-oblivious QRAC\cite{porac} exists with 
\be{poracp}
p=\frac{1+\frac{1}{\sqrt{n}}}{2}
\ee
and that this is the theoretical upper bound for parity-oblivious QRACs.  

We define a generalized parity-obliviousness as follows:  Let $I\subset \{0,...,n-1\}$ be a set of at least two $d$-level indices.  We define the $d$-parity of the corresponding set of $d$-levels by 
\be{}
P_I(a)=\sum_{i\in I} c_i(a_d)\mod d.
\ee

An encoding scheme is $d$-parity-oblivious iff there exists no index set $I$ and parity values $J, J'$ such that
\be{}
d^{1-n}\sum_{a|P_I(a)=J} \rho_a\neq d^{1-n}\sum_{a|P_I(a)=J'}.
\ee
$2$-parity is then the same as ordinary parity.  $d$-parity obliviousness was also introduced in \cite[dporac] shortly after the first version of this paper was made public.

\subsection{Parity-obliviousness of codes}
We now show that the encoding scheme in Eq. (\ref{racenco}) is $d$-parity-oblivious.  We let $I$ be an index set and $J\in \mathbb{Z}_d$ a parity value.  Now, 
\be{}
\begin{array}{l}
d^{1-n}\sum_{a|P_I(a)=J}\rho_a=\frac{d^{2-n-m}}{n}\sum_{i=1}^n\sum_{a|P_I(a)=J}\pi_{ic_i(a_d)}\\
=\frac{d^{2-n-m}}{n}\sum_{i=1}^n\sum_{j=0}^{d-1}\sum_{a|a_i=j, P_I(a)=J}\pi_{ij}\\ 
=\frac{d^{2-n-m}}{n}\sum_{i=1}^n\sum_{j=0}^{d-1}d^{n-2}\pi_{ij}=d^{-m}\id,\end{array}
\ee
where we used that every digit occurs equally often among inputs with a given parity, and that $\sum_{j=0}^{d-1} \pi_{ij}=\id$.   This shows that a mixed state describing a uniform distribution of all states with a specific $d$-parity is the maximally mixed state.  Since this state does not depend on the $d$-parity, the encoding scheme given by Eq. (\ref{racenco}) is $d$-parity-oblivious.  We also note that $d$-parity-obliviousness is conserved if we scale all the encoding Bloch vectors with a common factor, as we do when we place the encoding states on the insphere.  The sum of all Bloch vectors for the encoding states of inputs with a given parity will then still sum up to the $0$-vector, giving the maximally mixed state.  

We also note that no $d$-level value is special.  We may permute the values on the $d$-levels and still have $d$-parity-obliviousness.  This gives additional equations for the joint probabilities for $d>3$.

\subsection{Joint probabilities from parity-obliviousness}
\label{subsec:joint-prob}
We will now see that parity-obliviousness allows us to calculate some probabilities that we have neglected until now.  We focus on the $d=2$ case.  We know that a classical RAC encodes $n$ bits such that any bit can be retrieved correctly with a probability $p$.  Since a classical RAC involves no projective measurements, every bit can be obtained simultaneously, each with success probability $p$.  A $(4^m-1,m,p)$ QRAC allows one to retrieve $2^m-1$ bits,  as this is the number of PVMs that are diagonal in each of the MUB.  In a general setting, we may assume that we obtain $\nu$ bits in a parity-oblivious way, each with an individual success probability $p$.  An interesting quantity is then the probability $p(k,\nu)$, the probability that exactly $k$ of the obtained bits are correct.  This probability can be calculated if we assume uniform input.  Then, for $0<\nu'\leq\nu$, we have 
\be{po-subset}
p(k,\nu'-1)=\frac{\nu'-k}{\nu'}p(k,\nu')+\frac{k+1}{\nu'}p(k+1,\nu'),
\ee
since we may see the $\nu'-1$ bits as a random subset of $\nu'$ bits.  Parity-obliviousness implies that the probability for obtaining an odd number of correct bits is the same as the probability for obtaining an even number of correct bits when the number of bits is at least two.
\be{po-po}
\sum_{k=0}^{\nu'} (-1)^k p(k,\nu')=0,\ \nu'\geq 2
\ee
We know that 
\be{}
p(0,1)=1-p,\ p(1,1)=p.
\ee
If we assume that $p(k,\nu'-1)$ is known, then Eq. (\ref{po-subset}) gives $\nu'$ linearly independent equations for the $\nu'+1$ unknown probabilities $p(k,\nu')$.  Eq. (\ref{po-po}) gives the final equation, linearly independent from the others.  Trying to write it as a linear combination of the others leads to coefficients with alternating signs, but the coefficient for the $k=0$ equation must be positive, while the coefficient for the $k=\nu'-1$ equation must have a sign $(-1)^{\nu'}$, giving a contradiction.  The probabilities can now be calculated inductively, giving
\be{po-pof}
p(k,\nu')=2^{-\nu'}\binom{\nu'}{k}(1+(2k-\nu')(2p-1)).
\ee
Since $p(0,\nu)\geq 0$, we get the upper bound
\be{po-bound}
p\leq\frac{1+\frac{1}{\nu}}{2}.
\ee 

For classical RACs, $\nu=n$ and we see that our constructions give the optimal $p$ for parity-oblivious RACs.   This bound was also given in \cite{pcppom,porac}, but then only considering the strategy of encoding one bit perfectly, and guessing the remaining $n-1$ bits.   This gives the same average success probability as our classical RACs gives when guessing any bit.

The $(4^m-1,m,\frac{1+\frac{1}{(2^m-1)\sqrt{2^m+1}}}{2})$ QRACs do not reach the upper bound (\ref{po-bound}), but we see that the bias has a factor $\frac{1}{2^m-1}=\frac{1}{\nu}$.  The additional factor of $\frac{1}{\sqrt{2^m+1}}$ is the relative component size of the encoding Bloch vector in one of the $2^m+1$ orthogonal subspaces corresponding to the different MUB.   It is however not clear that multiplying these two constraining factors gives the optimal $p$.  We will now see that improvements are possible.

\section{Optimization}
\label{sec:opt}

We now discuss improvements of the QRACs.  The classical RACs are already optimized under parity-oblivious conditions.  We first look into the parity-oblivious possibilities for $d=2$, and divide into two cases, maximal and non-maximal $n$.  We then discuss other potential improvements, including $d>2$ and dropping parity-obliviousness.

\subsection{The worst case probability for maximal n}

We now consider $2$-level $(4^m-1,m,p)$ QRACs using the PVM operators of section \ref{subsec:good_two}.  Our goal is to scale up the encoding Bloch vectors to maximal length.  Eq. (\ref{poflambda}) gives the worst case probability in terms of an eigenvalue $\lambda$, where $-\lambda$ is the most negative eigenvalue among the eigenvalues of matrices on the form 
\be{}
\Sigma(\beta)=\sum_{k=1}^{4^m-1}(-1)^{\beta(k)}\bigotimes_{i=1}^m\sigma_{c_i(k)},
\ee
where $\beta$ can be any function $\beta:\{1,...,4^m-1\}\rightarrow\{0,1\}$.  The number of functions $\beta$ is $2^{4^m-1}$, so calculating the eigenvalues of all the possible matrices is very demanding already for $m\geq 3$.  We may however learn something from special cases.  The matrix

\be{produktsigma}
\Sigma(\beta:\beta(k)=-1\forall k)=\id_{2^m}-(\id+\sigma_x+\sigma_y+\sigma_z)^{\otimes m}
\ee
has $\binom{m}{k}$ eigenvalues that are $1-(\sqrt{3}+1)^m(\sqrt{3}-1)^{m-k}$, but most importantly, one eigenvalue which is $1-(1+\sqrt{3})^m$.  This gives a lower bound to $\lambda$, and thereby an upper bound to the worst case success probability.

\be{maximalnbound}
p\leq \frac{1+\frac{1}{(1+\sqrt{3})^m-1}}{2}
\ee
We could have replaced any of the tensor factors of $(\id+\sigma_x+\sigma_y+\sigma_z)^{\otimes m}$ in eq. (\ref{produktsigma}) with factors on the form $\id\pm_x\sigma_x\pm_y\sigma_y\pm_z\sigma_z$, and still obtained the same $\lambda$, meaning that at least $8^m$ encoding states are surface states on the Bloch sphere if we adjust a QRAC to have this success probability.  We have checked numerically that no $\Sigma(\beta)$ has an eigenvalue less than $1-(1+\sqrt(3))^m$ for $m=1,2$.  This means that we can obtain equality in the bound (\ref{maximalnbound}) in these cases.  We conjecture that this is the case for all $m$, i.e.:

\be{conjecture}
||\sum_{k=1}^{4^m-1}(-1)^{\beta(k)}\bigotimes_{i=1}^m\sigma_{c_i(k)}||\leq (\sqrt{3}+1)^m-1
\ee
implying that $p=\frac{1+\frac{1}{(1+\sqrt{3})^m-1}}{2}$.  Our attempts at proving this have not yet succeeded.

For $m>2$, the state space is too vast to cover with a numerical search.  We have nevertheless performed random numerical searches up to $m=6$ to look for $\Sigma(\beta)$ with an eigenvalue more negative than $1-(1+\sqrt{3})^m$.  The results are shown in table \ref{table:randomeigenvalues}, and no such eigenvalue was found.  The random searches do however not say much about the probability for such an eigenvalue to exist, since the number of checked matrices is much lower than the total number.  We can increase the number of excluded matrices by noticing that a sign flip on a term will not change an eigenvalue by more than $2$.  For instance, in the case $m=6$, we see that at least 139 signs must be flipped for an eigenvalue of any of the drawn matrices to break $-414.85$.  Because of this, each of the $10^5$ drawn matrices excludes more than $10^{262}$ other matrices too, but this is still only an unimaginably small portion of the total set of states.  We can not exclude the possibility of encoding states that break the conjecture (\ref{conjecture}), but the numerical searches show that it is unlikely to randomly stumble upon such a state.

\begin{table}
\begin{center}
\begin{tabular}{|l|c|c|c|c|}
\hline
$m$ & $\#\text{tries}$ & $\frac{\#\text{tries}}{\#\text{states}}$ &  $1-(1+\sqrt{3})^m$ & $-\lambda*$ \\ \hline
$2$ & $2^{15}$  & $1$  & -6.4641   & -6.4641 \\ \hline
$3$ & $25\cdot 10^{6}$  & $2.71\cdot 10^{-12}$  & -19.392  & -18.528 \\ \hline
$4$ & $4\cdot 10^{6}$  & $6.91\cdot 10^{-71}$  & -54.713  & -37.968 \\ \hline
$5$ & $10^{6}$  & $1.11\cdot 10^{-302}$  & -151.21   & -72.646 \\ \hline
$6$ & $10^{5}$  & $1.91\cdot 10^{-1228}$  &  -414.85  & -137.63 \\ \hline
\end{tabular}
\caption{\label{table:randomeigenvalues} Results of drawing $\#\text{tries}$ random $\Sigma(\beta)$ matrices. For $m=2$ all matrices have been checked, while for $m>2$ the number of tries is very small compared to the number of encoding states.  $-\lambda*$ is the most negative eigenvalue that is found and it is in all cases less negative than $1-(1+\sqrt{3})^m$.} 
\end{center}
\end{table}

\subsection{Parity-oblivious QRACs for m=2}
In the case of maximal $n$, all the orthogonal measurement directions are used, while some are omitted when a non-maximal number of bits is encoded.  Different subsets gives different worst case probabilities when we optimize the encoding states.  Finding the optimal subset is a complicated problem, but we have some clues.  The generators that we have used for $SU(2^m)$ have the property that every pair of generators either commute, or anticommute.  Moreover, the generators are self-inverse.  If $\{\sigma_i\}_{i=1}^n$  is a set of $n$ such generators that are all anticommuting, then 
\be{}
(\sum_{i=1}^n \pm_i \sigma_i)^2=n\id_{2^m}, 
\ee
and since $\sum_{i=1}^k \pm_i \sigma_i$ is traceless, its eigenvalues must be $\pm \sqrt{n}$ where the eigenvalues occur with equal multiplicity.    This means that we can obtain a worst case probability of 
\be{sqrtn-bias}
p=\frac{1+\frac{1}{\sqrt{n}}}{2}
\ee
as long as we can find $n$ anticommuting generators.  This is possible for $n\leq 2m+1$ and gives exactly the optimal parity-oblivious QRACs when there is no limitation on $m$\cite{porac}.   One such maximal set of generators is
\be{}
\{\sigma_x^{\otimes m}\}\cup\{\sigma_x^{\otimes k}\otimes(\sigma_y,\sigma_z)\otimes\id_{2^{m-k-1}}\}_{k=0}^{m-1}.
\ee
On the other hand, if $\{\sigma_i\}_{i=1}^n$  all commute, then we may find a set of signs $\{\pm_i\}_{i=1}^n$ such that $\sum_{i=1}^n \pm_i \sigma_i$ has an eigenvalue that is $-n$.  This suggests that the best subset of $\sigma$ matrices contains relatively few commuting pairs.

\begin{figure}
\includegraphics[scale=.6]{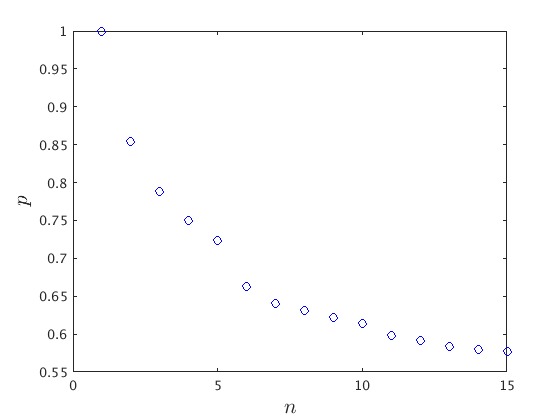}
\caption{\label{m2plot}Success probabilities for composite $d=2$, $m=2$ QRACs.}
\end{figure}

For $m=2$, we have tried every possible combination of $n$ PVM's to find the combinations that give the best worst case probability.  For $n=1$ to $n=5$, we get sets of anticommuting matrices, and a worst case probability according to eq. (\ref{sqrtn-bias}), the results for $n=6$ to $n=15$ is shown in table \ref{table:pformis2}.  For $n=6$ to $n=10$, the optimal subsets contain no triples of mutually commuting generators.  This means that only up to two bits can be recovered simultaneously.  Also, $-\lambda$ can be found from squaring the traceless part of $\Sigma^2$ in these cases.   For $n=11$ to $n=15$, there must always be some triples of bits can be obtained simultaneously.  We have plotted the success probability in figure \ref{m2plot}.  We see that $p$ drops significantly from $n=5$ to $n=6$ and from $n=10$ to $n=11$, indicating that the number of possible bits that can be obtained simultaneously has a significant impact on $p$.  This is not surprising since we have seen that simultaneous knowledge of several bits restricts $p$ under parity-obliviousness.

\begin{table}
\begin{center}
\begin{tabular}{|l|c|c|}
\hline
$n$ & $p$ & $p_{\text{insphere}}$ \\ \hline
$6$ & $\frac{1+\frac{1}{\sqrt{6+\sqrt{12}}}}{2}\approx 0.6625$  & $0.6179$  \\ \hline
$7$ & $\frac{1+\frac{1}{\sqrt{7+\sqrt{32}}}}{2}\approx 0.6405$  & $0.6091$  \\ \hline
$8$ & $\frac{1+\frac{1}{\sqrt{8+\sqrt{44}}}}{2}\approx0.6307$  & $0.6021$  \\ \hline
$9$ & $\frac{1+\frac{1}{\sqrt{17}}}{2}\approx0.6213$  & $0.5962$  \\ \hline
$10$ & $\frac{1+\frac{1}{\sqrt{10+\sqrt{84}}}}{2}\approx0.6142$  & $0.5913$ \\ \hline
$11$ & $\frac{1+\frac{1}{\sqrt{3}+\sqrt{2(4+\sqrt{3})}}}{2}\approx0.5977$  & $0.5870$ \\ \hline
$12$ & $0.5917$  & $0.5833$ \\ \hline
$13$ & $\frac{1+\frac{1}{\sqrt{7}+\sqrt{2(3+\sqrt{7})}}}{2}\approx0.5832$  & $0.5801$ \\ \hline
$14$ & $0.5800$  & $0.5772$ \\ \hline
$15$ & $\frac{1+\frac{1}{3+2\sqrt{3}}}{2}\approx 0.5774$  & $0.5745$ \\ \hline
\end{tabular}
\caption{\label{table:pformis2} Improved success probabilities compared to the insphere success probabilities.  The closed form expressions for $12$ and $14$ are omitted due to casus irreducibilis.} 
\end{center}
\end{table}

\subsection{Other parity-oblivious possibilities}

We have so far considered optimizing parity-oblivious QRACs for $d=2$.  We have performed numerical optimization for $d>2$ also, but the complexity of the problem limits the numerics.  For $d=4,\ n=4, m=1$ we find that encoding states can be improved such that
\be{}
p=\frac{1+\frac{1}{\sqrt{5}}}{4}=0.36180\rightarrow p=0.41350 
\ee
For $d=3,\ 5,\ 7$, $m=1$ and with maximal $n$, we find that some of the encoding states have singular density matrices, and therefore can not be improved.  In particular, we find that $d^2$ of the encoding states have $\frac{d-1}{2}$ eigenvalues that are $0$ and $\frac{d+1}{2}$ eigenvalues that are $\frac{2}{d+1}$.  This might well be true for all odd primes or even powers of an odd prime $d$, and may provide a clue to solving the problem of improving the encoding states in general.  It is not surprising if there is a distinction between even and odd primes, since the recipes for constructing MUB are different for even and odd primes\cite{mub}.

An attempt to find a solution could use other sets of mutually unbiased PVM operators than the ones we have used until now.  So far, we have used MUB in conjunction with mutually orthogonal arrays.  There are however some alternatives.  Firstly, it is not clear if every possible way to create MUB gives the same QRAC properties.  Maximal sets of MUB that are not unitarily equivalent exist\cite{inequi}.  Secondly, at least for $d=2$, we may make alterations to the PVMs that maintain mutually unbiasedness and projection valued measure properties, but mix up the commutation relations.  Our choice so far of generators of $SU(2^m)$ have the property that each pair either commute or anticommute.  If we choose a subset of generators, $\{\sigma_i\}_{i=1}^k$ that all anticommute, then we may create a new set $\{\sigma_i'=\sum_{j=1}^k O_{ij}\sigma_j\}_{i=1}^k$, where $O_{ij}$ is a rotation matrix.  The new set still contains anticommuting matrices and the eigenvalues are unchanged.  If we put them together with the rest of the generators, then the generators still give a maximal set of mutually unbiased PVMs, but the unaltered and altered generators will no longer always either commute or anticommute.   

It remains to investigate this direction thoroughly.  As an initial test, for $d=m=2,\ n=15$, we have grouped together five triples of mutually anticommuting generators and applied random rotation matrices to each triple.  This gave a worse worst case success probability in all tests, but we cannot say that a systematic scheme will not improve the probability instead.  However, the idea may be more fruitful for non-maximal $n$, where the encoding scheme is less symmetric.  For $n=15$, every measure operator commutes with 6 other operators, while for $n=10$, every operator commutes with 3 other operators, but for other values larger than $n=5$, this symmetry is not present, and this may encourage alterations to the measure operators.

\subsection{Dropping parity-obliviousness}

More possibilities arise if we neglect parity-obliviousness.  For instance, a $(6,2, \frac{1+\frac{1}{\sqrt{3}}}{2})$ QRAC can be constructed as two $(3,1, \frac{1+\frac{1}{\sqrt{3}}}{2})$ QRACs.  Adding together smaller QRACs in this way may also favor using QRACs with rectangular encoding schemes, i.e. QRACs with varying bloch component sizes in the different measurement directions. A $(4,3,0.898)$ arises from constructing three rectangular 1-qubit QRACs, encoding the three first bits into separate qubits with a success probability $0.898$, and encoding the fourth bit into each qubit with a success probability $0.802$.  The probability for success in two or three out of three when measuring the fourth bit is then also $0.898$.  This type of composite QRACs requires more general solutions than we have presented so far and may be subject for future research.

Composite QRACs can only be made if $n$ is small enough to be covered by two or more separate QRACs sharing the total number of q-bits or q-$d$-levels $m$.  But non-parity-oblivious codes can also improve the worst case success probability in other ways.  The numerical solutions that we obtained for $d=2$, $m=2$ with only pure states beat the mixed states solution for low values of $n$, as we see in table \ref{table:2}.  The cases $n=7,8$ do not allow composite QRACs, so there must be a different scheme that optimizes these cases.


The parity-oblivious hypercube solutions use encoding states that only span $n$ out of the $2^{2n}-1$ dimensions of Bloch space, while the encoding states of the $(3m,m,\frac{1+\frac{1}{\sqrt{3}}}{2})$ QRAC that consists of $m$ cubic QRACs do however span the whole of Bloch space.  It is therefore not surprising if non-parity-oblivious solutions give better worst case probabilities than the hypercube solutions for low values of $n$, as they may take advantage of more directions in Bloch space.  On the other hand, the maximal $n$ hypercube solution uses all the state space dimensions, and it seems that parity-obliviousness and optimal worst case probability coincide in this case.

\begin{table}
\begin{center}
\begin{tabular}{|l|c|c|c|c|c|c|}
\hline
$n$ & 7 & 8 & 9 & 10 & 11 & 12\\ \hline
$p$ & 0.68412  & 0.65249  &  0.60319  & 0.53919  & 0.52468 & 0.50054 \\ \hline
$\bar{p}$ & 0.72839 & 0.71653  & 0.70268 & 0.66544  & 0.66177 & 0.65562\\ \hline
$p_{\mbox{mix}}$ & 0.6405 & 0.6307 & 0.6213 & 0.6142 & 0.5977  & 0.5917 \\\hline
\end{tabular}
\caption{\label{table:2}Numerically obtained $(n,2,p)$ QRACs with average success probabilities $\bar{p}$ and analytical $(n,2,p_{\mbox{mix}})$ QRACs.} 
\end{center}
\end{table}

\section{RACs vs QRACs}
\label{sec:qvsc}

A recurring theme in quantum information theory is the correspondence between two bits and one qubit, or in general between $m$ quantum $d$-levels and $2m$ classical $d$-levels.  This occurs because the dimensions of the state spaces coincide.  It has been shown\cite{qic} that $m$ qubits can be used to send at most $2m$ bits of information, using previously shared resources.  The super dense coding protocol\cite{sdc} achieves this, using $m$ maximally entangled pairs.  Since additional resources are necessary in order to use one qubit to transmit two bits, one may be lead to believe that the two bits contain more information than a qubit.  However, an entangled pair is also necessary in order to send a qubit using two bits.  This means that the two different information carriers both have advantages over each other.  This is also the case in the setting of (Q)RACs.  We will restrict the discussion to powers of a prime $d$, since these are the cases where we have some understanding of the optimized solutions.  

In the most basic example, 1 qubit vs 2 bits, we have that both can encode 1 bit faithfully.  Sending 2 bits with a qubit only allows the receiver to get 1 of the bits with a success probability of $\frac{1+\frac{1}{\sqrt{2}}}{2}$, while 2 bits of course can send 2 bits faithfully.  The $(3,1,\frac{1+\frac{1}{\sqrt{3}}}{2})$ QRAC do however beat the $(3,2,\frac{2}{3})$ RAC.   

We saw in section \ref{subsec:joint-prob} that for $d=2$, with parity-oblivious (Q)RACs, the number of bits $\nu$ that can be retrieved simultaneously restricts the success probability.  This gives a fundamental understanding of the differences between an $(4^m-1,2m,p)$ RAC and an $(4^m-1,m,p')$ QRAC.  The RAC allows the receiver to recover information about every bit, as opposed to the QRAC, where a subset of bits must be chosen, while all information about the rest of the bits is erased following the measurement.  This loss of information is however necessary for the QRAC to give a better success probability than the RAC.  This is if we assume parity-obliviousness, but this is a consequence of orthogonal measurement directions in Bloch space, which seems hard to do without if we want an optimal worst case (Q)RAC for maximal $n$. The success probabilities for some bits will otherwise depend on the value of other bits.

The QRACs we have presented allow the receiver to obtain information about a subset of up to $\frac{d^m-1}{d-1}$ of the $n\leq \frac{d^{2m}-1}{d-1}$ $d$-levels, but we can also construct a POVM that obtains information about every $d$-level, making the QRAC similar to a classical RAC.  This is done by using measure operators that are proportional to the encoding states.
\be{cracmeasure}
F_a=d^{m-n}\rho_a, \ \sum_{a} F_a=\id.
\ee

The probability for measuring the $i$'th $d$-level to the correct value $a_i$ is then 
\be{}
p=\Tr (\rho_a\sum_{b|b_i=a_i}d^{m-n}\rho_b)=\frac{1}{d}+\frac{1}{2n}r_a^2,
\ee
where $r_a$ is the Bloch vector length of the states $\rho_a$.  The maximal value $r_a=\sqrt{2(d-1)d^{-m}}$ occurs when $\rho_a$ has $d^{m-1}$ non-zero eigenvalues that are all $d^{1-m}$.   In this case we retrieve the classical RAC probability $p=\frac{1+\frac{d-1}{n}}{d}$.  In general, the encoded message of an $(n,m,p)$ QRAC can be interpreted to give a success probability 
\be{}
p_{q\rightarrow c}=\frac{1+\frac{(dp-1)^2}{d-1}}{d}
\ee  
such that the same information about every $d$-level is obtained.  

For $d=2$ QRACs, $p_{q\rightarrow c}$ is coinciding with the classical RAC probability for up to $n=2m+1$ bits.  This coincides with the range where optimal $p=\frac{1+\frac{1}{\sqrt{n}}}{2}$ parity-oblivious QRACs are availlable.  For large values of $n$, the QRAC will give a higher $p$ than the classical RAC, which again gives a higher success probability than the QRAC measured with the measure (\ref{cracmeasure}).  For instance, encoding 15 bits in 2 qubits gives $p=0.5774$, with $p_{q\rightarrow c}=0.5120$, while the classical RAC using 4 bits has $p=0.5333$.  \\



\section{Discussion and outlook}
 \label{sec:concl}

We have explained Bloch space geometry and used it to construct a previously known class of mixed state QRACs that encode the maximal number of bits ($n=4^m-1$) into $m$ qubits in such a way that any single bit can be recovered with a probability greater than $\half$. The previously known codes use projection valued measures that are orthogonal in Bloch space, both of which seem necessary for an optimal code.  We have improved the encoding states of the codes to give a higher worst case success probability.  We also saw that up to $n=2^m-1$ bits can be encoded into m classical bits under the same worst case requirements.  This means that $2m$ bits and $m$ qubits can encode the same number of bits.  The correspondence between 1 qubit and 2 classical bits is frequently seen and this is because the dimensions of state spaces coincide.  The classical and quantum random access codes both exhibit advantages over the other.  We have found that the probability of success is higher for the quantum case, but in the classical case, there is no projective measurement and we do not have to choose which bits to obtain information about.  We have seen that these (Q)RAC constructions are parity-oblivious, and that this implies that the worst case success probability is at most $\frac{1+\frac{1}{\nu}}{2}$, where $\nu$ is the number of bits one can obtain information about simultaneously.  The loss of information during a projective measurement is therefore a necessity for the QRAC to outperform the RAC in terms of worst case success probability.  We also saw that one can obtain information about every bit from the QRAC too, but this gives a worse success probability than for the classical RAC for $n>2m+1$.

We have generalized the problem to a situation where $n$ classical $d$-levels are encoded in $m$ (quantum) $d$-levels such that the worst case probability for any wrong outcome when decoding one $d$-level is less than $\frac{1}{d}$.  The correspondence between classical RACs and QRACs is also seen for $d$-levels.  The solution for $d=2$ generalizes via mutually unbiased bases in combination with mutually orthogonal arrays.

For the classical RACs, we have achieved the optimal $d$-parity-oblivious success probability $p=\frac{1+\frac{d-1}{n}}{d}$, while for QRACs, the question of optimality is harder.  For $d=2$, we have seen that improved encoding states allow an improved success probability over what was previously known.  For $n\leq2m+1$, the optimal solution is already known, while for maximal $n=4^m-1$, we have conjectured that $p=\frac{1+\frac{1}{(1+\sqrt{3})^m-1}}{2}$ and tested that it holds for $m<3$.  The proof or disproof of this depends on an eigenvalue problem which has an appealingly simple description.  

For $2m+1<n<4^m-1$, the optimization is less straight forward for general $m$.  However, for $m=2$, we have optimized the subset of measure operators, and the results are candidates for optimal parity-oblivious QRACs. 
The low $n$ solutions are however not optimal when non-parity-oblivious codes are allowed, but it seems that the optimal solution for maximal $n$ is parity-oblivious too.  The parity-oblivious hypercube solutions make use of $n$ dimensions in Bloch space.  How to utilize all of the dimensions systematically for low values of $n$ is an interesting question that remains to be answered. 
  
For $d>2$, we have found for the solutions that are within exhaustive search distance that the encoding states of our solutions can not be improved upon in the maximal $n$ case.  It still remains to be shown if this is the case in general or not.  Also, as in the case for $d=2$, the non-maximal $n$ cases may be even harder to solve.  It would be interesting to see if an optimal $d$-parity-oblivious QRAC that generalizes \cite{porac} can be found.  This may however be difficult, since for $d>2$, the traceless parts of the measure operators do not always either commute or anti-commute.



\section*{Acknowledgements}
\noindent
The author acknowledges support from grant NFR213606 and thanks Olav Sylju\aa sen for comments on the written manuscript.
\




\end{document}